\documentclass[a4paper,10pt,onecolumn,nofootinbib]{article}
\usepackage[latin1]{inputenc}
\usepackage{amsmath}
\usepackage{amsfonts}
\usepackage{amssymb,amsthm}
\usepackage{graphicx}
\usepackage[left=2cm, right=2cm, top=2.5 cm, bottom=2.5 cm]{geometry}
\usepackage{float}
\usepackage{graphicx}
\usepackage{hyperref}
\usepackage[english]{babel}
\usepackage{subcaption}
\usepackage[justification=centering,singlelinecheck=false]{subcaption}
\usepackage[justification=RaggedRight,singlelinecheck=false]{caption}
\usepackage{amsfonts}
\usepackage{braket}
\usepackage{mathrsfs}
\usepackage{empheq}
\usepackage[shortlabels]{enumitem}
\usepackage{xcolor}
\usepackage{pgfplots}
\usepackage{bigints}
\usepackage{setspace}

\usepackage{graphicx}
\usepackage{dcolumn}
\usepackage{bm}

\title{\textbf{The $\rho$ parameter and the CDF - $II$ $W$-mass anomaly: observations on the role of scalar triplets}}
\author{Rituparna Ghosh, Biswarup Mukhopadhyaya, Utpal Sarkar 
\footnote{\href{mailto:rg20rs072@iiserkol.ac.in}{rg20rs072@iiserkol.ac.in} \href{mailto:biswarup@iiserkol.ac.in}{biswarup@iiserkol.ac.in} \href{mailto:utpal.sarkar@iiserkol.ac.in}{utpal.sarkar@iiserkol.ac.in}}\\
{Department of Physical Sciences} \\ {Indian Institute of Science Education and Research, Kolkata, Mohanpur - 741246, India}}
\date{}

\begin{document}

\maketitle

\begin{abstract}
The $\rho$-parameter, together with the W-and Z-masses, acts as Occam's razor
on extensions of the electroweak symmetry breaking sectors.
We apply this to non-doublet Higgs scenarios, by examining
 the CDF- $II$ claim on the W-boson mass.  Suspending any judgement on the
 CDF claim, we show that
in general, if one works at the tree level, theoretical models which predict $\rho =1$ at the tree-level
are  inconsistent with the CDF claims at 4-6 standard deviations
 if one confines oneself to  the existing Z-boson mass, and  the earlier $M_W$ value 
 from either the global fit or the ATLAS data.
We take some well-motivated scenarios containing one or more scalar SU(2) triplets in addition
to the usual doublet, and show  that, both a scenario including a
complex scalar triplet and one with a complex as well as a real triplet (the Georgi-Machacek model)
can be made consistent with the new data, where a small splitting between the complex and the real triplet vacuum expectation values is required in the second scenario. We will explore the consequences of this splitting, either at tree level or via incalculable new physics contribution to $M_W$, and indicate, as illustrations its implications in $H^{\pm}W^{\mp}Z$ type interaction vertices. 

\end{abstract}

\section{Introduction}
A pertinent question facing all of us today is: is electroweak symmetry breaking(EWSB) entirely driven by one or more scalar doublets or can scalars in other $SU(2)$ representations {,too,} have a role there? Since the symmetry breaking process is inevitably reflected in the weak gauge boson masses and their phenomenology, any new information of the W and Z boson masses gets linked with the above question.
\newline 
Of considerable relevance in this context  is the recently announced estimate of the W-boson mass by the CDF-$II$ collaboration
setting the value at $M_W = 80,433.5 \pm 6.4_{Stat} \pm 6.9_{Sys}$ MeV \cite{ref1}, which is in apparent disagreement with { the standard model (SM) expectation of $M_W = 80,357 \pm 4_{inputs} \pm 4_{theory}$ MeV \cite{ref30} } at the $7.2 \sigma$   level. While the claim is quite justifiably subject to closer scrutiny, it serves as a stimulant to theoretical introspection, keeping especially in mind the
fact that it involves the EWSB sector. This sector
associates a number of important issues, such as the role of additional fields
in EWSB, the generation of neutrino Majorana masses, and hence the possibility of
lepton number violating interactions.
One thus feels inclined to forge options beyond the standard model (BSM) on the anvil of the recently published data, just to prepare ourselves for the
eventuality that the announced disagreement persists. Several speculations in this direction have already appeared.\cite{ref2}- \cite{ref23}

A rather curious feature of the experimental results is that the Z-boson mass  stands at the 
value  measured rather precisely at the Large Electron-Positron (LEP) collider,
namely, $M_Z = 91.1876 \pm 0.0021$GeV \cite{ref24}. This generates an apparent tension with the value of
the $\rho$-parameter, defined as $\rho = m^2_W/(m^2_Z \cos^2 \theta_W)$ ($\theta_W$ being the weak mixing angle),
whose tree-level value in the standard model (SM) is unity, and the experimentally identified
range is $1.00038 \pm 0.0002$. The tension arises essentially because both of 
the gauge boson masses owe themselves
to the same `effective' Higgs vacuum expectation value (vev) $v$, namely,
$246.22/\sqrt{2}$ GeV. On the other hand, any
BSM option in EWSB, too, has a bearing, amongst other things,  on the $\rho$-parameter. 
Many such options also affect different aspects of electroweak phenomenology.

While the jury is still out on the CDF claim, we consider it important as an interim
step to connect it to already proposed extensions of the EWSB sector. An example is a class of 
models containing one or more scalar triplets in addition to the SM doublet. Over and above the neutrino mass generation, they have additional rich phenomenology
in store, including  the possibility of substantial non-doublet contributions to EWSB and the associated signal \cite{Godbole:1995sx,Ghosh:1996jg,Chakrabarti:1998qy}. It is naturally of interest to see what kind of additional constraint comes on them
from the recent CDF claim. Such constraints, in which the $\rho$-parameter plays an important role,
are found by us, using earlier reference values of $M_W$ based on different sets of data. Thus, different sets of constraints appear on the triplet model parameter space, which we
report below. 
\newline 
The `new' conclusions we have arrived at are:

\begin{enumerate}
    \item If one works at tree level and uses the presently accepted  Z-mass, the W-mass as per  CDF claims,
 and the `old'  W-mass following  the global fit, then one has an inconsistency at the
level of $6.78 \sigma$ with any theoretical scenario which predicts $\rho = 1$ at the tree-level.
\item { As the simplest model deviating from the requirement of tree-level $\rho = 1$,
a scenario including a single $Y = 2$ scalar triplet  is consistent with the CDF-II claim once the oblique corrections are considered.
The triplet vacuum expectation value (vev) remains as constrained as before. }

\item The Georgi-Machacek scenario, comprising a complex ($\chi$) and a real ($\xi$) scalar triplet,
the triplet vevs can make substantial contributions to the W-and Z-masses so long as the vevs are equal, namely, $v_{\chi} = v_{\xi}$. However, such equality is inconsistent
unless one uses the W-mass based {\em only} on LEP and Tevatron data, and not one from
a global fit.  On the other hand, a region of the parameter space with $v_{\chi} \ne v_{\xi}$ is allowed,
which we identify below.
\end{enumerate}
In section 2, we present a general discussion on the (in)consistency of the recent data with
general scenarios that lead to $\rho = 1$ at the tree-level. Section 3 contains analyses of two 
kinds of triplet models. The first of one, namely, the Georgi-Machacek model,
has a complex and a real triplet, where $\rho = 1$ is restored via a custodial SU(2). The second of this is the model with a single complex triplet, which tends to
shift the tree-level value of $\rho$ from unity. The constraints
arising on each scenario from the recently reported value of $M_W$ are presented for each case.
We conclude in section 4. 

\section{Gauge Bosons mass}
The new result can be explained in two ways. One can enhance the mass of W boson by including loop effects as suggested \cite{ref10}\cite{ref26,ref27}. Another way of looking at is to accept that the scalar vev differs by a small amount\cite{ref28} from what is indicated by earlier measurements, in such 
a way that it remains consistent with the mass of the Z boson as measured at LEP\cite{ref24}. We go by the latter approach,
 demanding consistency with independently measured quantities like the mass of the Z boson and the 
 $\rho$-parameter.

In a general model consisting of more than one Higgs-like multiplets, the masses of the  W and Z bosons are given by\cite{ref29},

\begin{equation}
      M_W^2 = \frac{g^2}{4}{\Sigma_i (4T_i(T_i +1) - Y_i^2)c_i v_i^2} = \frac{g^2}{4}v_W^2
\end{equation}    
\begin{equation}
      M_Z^2 = \frac{g^2 + {g'}^2}{4}\Sigma_i 2Y_i^2 v_i^2 = \frac{g^2 + {g'}^2}{4} v_Z^2
\end{equation}

$c_i = 1/2$ for $Y = 0$ multiplet and $c_i = 1$ for $Y \ne 0$ multiplet. $T_i$ and $Y_i$ is the isospin and hypercharge of the $i^{th}$ higgs multiplet. $V_W$ and $V_Z$ denotes the contribution of Higgs vev to the mass of the W and Z boson respectively. Note that the added contribution from a triplet serves to enhance both $M_W$
and $M_Z$, But the value of the $\rho$-parameter is reduced in the process, as can be seen below.

In general, the $\rho$ parameter is given by\cite{ref29},
\begin{equation}
    \rho = \frac{{\Sigma_i (4T_i(T_i +1) - Y_i^2)c_i v_i^2}}{\Sigma_i 2Y_i^2 v_i^2}
\end{equation}
$\rho = 1$ at tree-level requires  $v_W^2 = v_Z^2 = v^2$ where $v \approx 246.22$ GeV . Now, if one 
goes by the recent  CDF claim, namely,  $M_W = 80.4335 \pm 0.0094$ GeV, let us concede that
$v$ is changed by a small amount \footnote{We have attributed the change in W-mass
to the driving vev, keeping the gauge coupling unchanged, which causes a lot less tension with weak universality etc}, so, the W mass is now given by
\begin{equation}
      M_W^2 =  \frac{g^2}{4}(v^2 + \Delta v^2) 
\end{equation}
\begin{equation}
    M_W = \frac{gv}{2}(1 + \frac{\Delta v^2}{2 v^2}) =  M_{W,old} + \Delta M_W 
\end{equation}
where
\begin{equation}
    \Delta M_W = \frac{g}{2}\frac{\Delta v^2}{2v} = \frac{\Delta v^2}{2v^2}{M_{W,old}} 
\end{equation}
Here $M_W$ is the W mass recently announced by CDF, $M_{W,old}$ is the old measured W-mass and V is the `effective' vacuum expectation value  which is required to reproduce that mass. Hence $\Delta v^2 $ is given by
\begin{equation}
   \frac{ \Delta v^2}{2 v^2}= \frac{\Delta M_W}{M_{W,old}}
\end{equation}
Now as there is no new measurement for the mass of the Z boson, this change in W mass should be consistent with the old measurement of the Z boson mass. For  models which naturally keep $\rho = 1$ at tree level, the altered `effective' vev should contribute  to $M_Z$ with similar functional dependence as what occurs in the
expression for $M_W$ . The thus modified mass of the Z boson is given by
\begin{equation}
      M_Z^2 = \frac{g^2 + {g'}^2}{4} (v^2 + \Delta v^2)  \ \ \
\end{equation}
Hence,
\begin{equation}
      M_Z = M_{Z,mean} + \Delta M_Z 
\end{equation}
where, $M_{Z,mean}$ is the central value of the old measured Z mass, namely,
91.1876 GeV. The consequent shift in $m_Z$ required to retain $\rho = 1$ at tree-level is denoted 
here by $\Delta M_Z$, and is given by
\begin{eqnarray}
    \label{10}
    \Delta M_Z &=& \frac{\sqrt{g^2 + {g'}^2}}{2}\frac{\Delta v^2}{2v}  \nonumber \\
    &=& \frac{\Delta v^2}{2v^2}M_{Z,mean} \nonumber \\  
    &=& \frac{M_{Z,mean}}{M_{W,old}}\Delta M_W
\end{eqnarray}

The existing measurements \cite{ref24} tell us that { $M_Z = 91.1876 \pm 0.0021$ GeV.} Table-1 then implies that for $M_{W,old}$ standardised following the world average\cite{ref30} and ATLAS measurements\cite{ref31} respectively, the demand that $\rho = 1$ shifts $M_Z$ to values disagreeing with $M_{Z,mean}$ at the 4.23$\sigma$ and 6.78$\sigma$ level respectively. For $M_{W,old}$ given by the LEP2 and Tevatron combined results, shifts $M_Z$ by less than 1$\sigma$ from $M_{Z,mean}$. Thus the first two choices of $M_{W,old}$ make the data clearly incompatible with any theoretical scenario that predicts $\rho = 1$ at the tree level. We emphasize that such incompatibility occurs even when $M_{W,old}$ is set at its maximum value at $2\sigma$ level (with $M_{Z}$ appropriately adjusted within
its  $2\sigma$ allowed range), and $M_{W,new}$, at its  minimum  $2\sigma$ value. The possibilities are summarised in Table-1

\begin{table}
\centering
\begin{tabular}{ |c|c|c| } 
 \hline
 $M_{W,old}$ & Shift from $M_{Z,mean}$  \\
 \hline
 \hline
 ATLAS & 4.23$\sigma$  \\ 
 \hline
 World Avg & 6.78$\sigma$ \\
 \hline
 LEP2/Tevatron &  $<$ 1$\sigma$  \\ 
 \hline
\end{tabular}
\caption{\centering The shift from $M_{Z,mean}$ for the various cases mentioned in the text}
\end{table}


\section{Model with Higgs triplet(s)}
We turn next to an illustrated class of BSM scenarios, where one or more $SU(2)$ triplet scalars participate in EWSB. Such scenarios are particularly interesting in the context of the Type-II seesaw mechanism of neutrino mass generation. In some variants, which we examine below the triplet vevs may also contribute substantially to the weak gauge boson masses, and thus play important roles in weak-scale collider phenomenology.

\subsection{A real and a complex triplet: the Georgi- Machacek (GM) model}
We take up first a scenario \cite{ref35,ref36} where one has a $Y=2$ complex triplet $\chi = (\chi^{++}, \chi^+, \chi^0)$ and a $Y=0$ real triplet $\xi = (\xi^+, \xi^0, \xi^-)$. As discussed widely in the literature,  $\langle \chi^0 \rangle = \langle \xi^0 \rangle = v_\chi$, ensured via a custodial global $SU(2)$ \cite{ref37} retains $\rho = 1$ at the tree-level. The scalar potential in this case is,
\begin{eqnarray}
\label{11}
					V(\Phi,X)&=&\frac{\mu_2^2}{2} Tr(\Phi^\dagger\Phi) + \frac{\mu_3^2}{2}       Tr(X^\dagger X) + \lambda_1[Tr(\Phi^\dagger\Phi)]^2 + \lambda_2Tr(\Phi^\dagger\Phi)Tr(X^\dagger X) \nonumber \\
					&+& \lambda_3Tr(X^\dagger X X^\dagger X)
					+\lambda_4[Tr(X^\dagger X)]^2 - \lambda_5Tr(\Phi^\dagger \tau^a \Phi \tau^b)Tr(X^\dagger t^a X t^b) \nonumber \\ 
					&-& M_1Tr(\Phi^\dagger \tau^a \Phi \tau^b)(UXU^\dagger)_{ab} - M_2Tr(X^\dagger t^a X t^b)(UXU^\dagger)_{ab}
\end{eqnarray}
with, 
	\begin{equation}
		\Phi = \begin{pmatrix}
		\phi^{0 \star} & \phi^+\\
		\phi^- & \phi^0 
		\end{pmatrix} \ \ \
		X = \begin{pmatrix}
						\chi^{0 \star} & \xi^+ & \chi^{++}\\
						\chi^- & \xi^0 & \chi^+\\
						\chi^{--} & \xi^- & \chi^0\\
					\end{pmatrix} \ \ \ 
		U = \begin{pmatrix}
    -\frac{1}{\sqrt2} & 0 & \frac{1}{\sqrt2}\\
    -\frac{i}{\sqrt2} & 0 & -\frac{i}{\sqrt2}\\
    0 & 1 & 0\\
    \end{pmatrix}
     \end{equation}

\begin{figure}[!htb]
         \centering
         \includegraphics[width=0.6\textwidth]{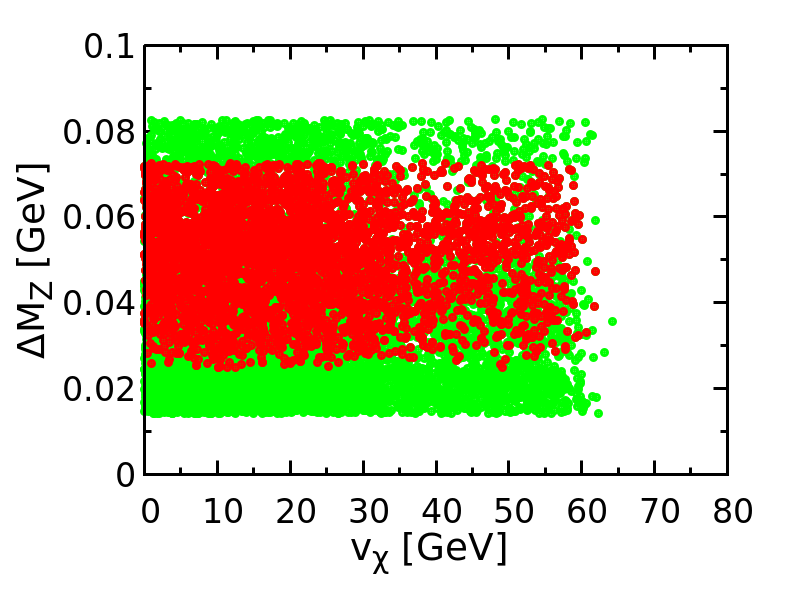}
         \caption{ The $2\sigma$ region consistent with the CDF claim is shown in green and the region consistent at $1\sigma$ level is shown in red. }
         \label{1ref}
\end{figure}

$\tau^a$ s and $t^a$ s are the $2\times2$ and $3\times3$ $SU(2)$ generators respectively.\newline
{If one wants to satisfy the CDF claim on $M_W$ without any new physics coming in, the resulting shift in $M_Z$ is shown in Fig. \ref{1ref}. Here $M_{W,old}$ was in accordance with the world average}

Equation \eqref{11} manifestly ensures $v_{\chi} = v_{\xi}$. However, as has been emphasised in the literature \cite{Gunion:1990dt, ref36}, it corresponds to a rather fine-tuned situation. The custodial $SU(2)$ is broken by the $U(1)_Y$ gauge interaction of the scalar triplets; quadratic divergences consequently crop up in higher order corrections to the weak gauge boson masses and the $\rho$-parameter. 
The cancellation of these divergences requires the introduction of counterterms which essentially depend on the UV-completion of the theory, with its accompanying new physics. The parameters in the GM scalar potential are thus subject to corrections that cannot be calculated within the model itself. One may at best use phenomenological values of these parameters. If the GM scenario has to be relevant in TeV-scale physics, then the loop-revised  mass parameters and the UV limit of the theory need to be around that scale. This is an assumption underlying any model-building, where observable consequences are expected.

We, too, take such a phenomenological approach and express the corrected W and Z masses, following \cite{Gunion:1990dt} as,

\begin{eqnarray}
M^2_Z = {\frac{1}{4}} (g^2 + g'^2)v^2 \\
M^2_W = {\frac{1}{4}} g^2 v^2 (1 + s^2_H \delta + {\frac{1}{2}} s^2_H \delta^2)
\end{eqnarray}

\noindent
where $v^2 = v_\phi^2 + 8v_\chi^2$; $s_H = \frac{2\sqrt{2}v_\chi}{v}$. {It should be emphasised that, $\delta$ is not effecting $M_Z$ as it is a measure of the separation of the real triplet vev from that of the complex triplet, i.e, $v_\xi = v_\chi(1+\delta)$.We take $v_\chi = v_\xi$ at the tree level at this stage. Note that $M_Z$ is fixed at its measured value, and the scalar vevs are regulated accordingly, while the effect of higher oder
corrections is explicitly shown in the expression for $M_W$.}

The quantity $\delta$ can be expressed as

\begin{equation}
\delta = \delta_0 + \delta_{div}
\end{equation}

\noindent
Where $\delta_{div}$ is the counterterm that cancels the quadratic divergence. Following the argument given above, the finite correction $\delta_0$ is not calculable without knowledge of the UV completion, and has to be fixed phenomenologically.

\begin{figure}[!htb]
         \centering
         \includegraphics[width=0.6\textwidth]{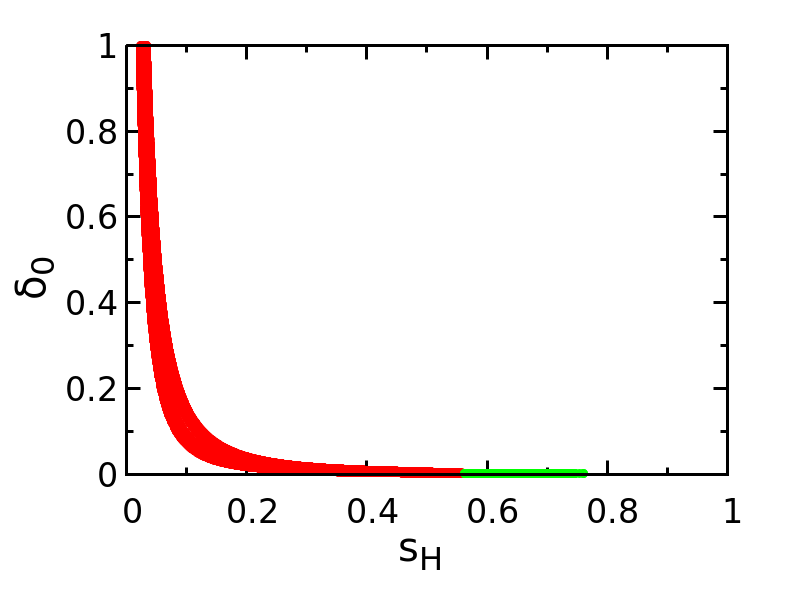}
         \caption{ The $2\sigma$ region consistent with both the CDF claim and $\rho$ parameter shown in Red and the region consistent with the oblique corrections in 2$\sigma$ level is shown in green. }
         \label{1f}
     \end{figure}

If now one tries to explain the CDF claim (say, at the $2\sigma$ level), one obtains a range of $\delta_0$ depending on the value of $s_H$ which is a measure of the triplet contribution to the weak boson masses. The thus allowed region in the $s_H - \delta_0$ plane, consistent at the $2\sigma$ level with the measured value of $\rho$ as well as the CDF claim, is shown in Fig.  \ref{1f}. Here one has to use the following expression for the $\rho$ parameter corrected by $\delta_0$
\begin{equation}
\rho = \frac{v_\phi^2 + 4v_\chi^2 + 4v_\chi^2(1 + \delta_0)^2}{v_\phi^2 + 8v_\chi^2}
\end{equation}
{ The free parameters for our analysis were, $\lambda_2,\lambda_3,\lambda_4,\Tilde{M_2},M_5^{++},M_3^{+},s_H$ and the corresponding ranges are shown in Table-\ref{t2}. $\lambda_1$ is tuned such that one of the two custodial singlet CP-even Higgs masses is always fixed at 125 GeV\cite{ref38}. $M_5^{++}$ and $M_3^+$ are the masses of the fiveplet state and three-plet state respectively and given by,
\begin{eqnarray}
      {M^{\pm\pm}_5}^2 &=& \frac{M_1}{4v_\chi}v_\phi^2 + 12M_2v_\chi + \frac{3}{2}\lambda_5v_\phi^2 + 8\lambda_3v_\chi^2 \\
    {M_3^{\pm}}^2 &=& (\frac{M_1}{4v_\chi} + \frac{\lambda_5}{2})v^2
\end{eqnarray}
}
The following features  of the GM scenario, implied by the CDF claim emerge from Fig.  \ref{1f}:

\begin{table}

\centering
\begin{tabular}{ |c|c|c| }
 \hline
$ {\lambda_2}$ & $[0,0.8] $ \\
 \hline
$ {\lambda_3}$ & $[-1.5,1.5] $ \\ 
 \hline
 $\tilde{\lambda_4}$ & [-1.5,1.5]  \\
 \hline
 $\tilde{M_2}$ & 10 GeV  \\ 
 \hline
 $M_5^{++}$ & 200-1000 GeV  \\
 \hline
 $M_3^{+}$ & $> M_5^{++}$   \\ 
 \hline
 $s_H$ & [0,1] \\ 
 \hline
\end{tabular}
\caption{\centering  Values of various parameters used in our scan to get Fig. \ref{1f} and \ref{2f}}
\label{t2}
\end{table}
\begin{enumerate}
\item The larger is $s_{H}$ the more is the value of $\delta_0$ restricted. It should be noted that limits from the LHC data allow $s_H$ to be upto 0.3-0.4 \cite{ref38, rb2022} approximately, depending on details of the GM parameter space. Thus, one gets restricted to $\delta_0$ below $0.1$ if the weak gauge bosons have to receive substantial contributions to their masses from triplet vevs. 
\item On the other hand, one is constrained to relatively small values of $s_H (\leq 0.1)$ in order to allow $\delta_0 \geq 0.5$. A large finite correction to $M_W$ arises presumably from some strongly coupled UV completion, thus implies a small contribution of triplet vev to the W- and Z- boson masses.

\item { Consistency with the observed values of oblique electroweak parameters is another question. In terms of oblique parameters $S,T$ and $U$\cite{ref34}, the shift in $M_W$ is given as,
\begin{eqnarray}
    \Delta M_W = \frac{\alpha}{2(c_W^2 - s_W^2)}(-\frac{1}{2}S + c_W^2 T + \frac{c_W^2 - s_W^2}{4s_W^2}U)
\end{eqnarray}
As seen from Fig. \ref{1f}, the region in green is consistent with oblique corrections though these values of triplet vev is highly disfavoured by collider data\cite{ref38,rb2022}.We have used reference\cite{indir} to obtain the oblique parameters for our case and our results are also consistent with reference \cite{referee1} for its chosen value of T-parameter.} If the range of $T$ parameter is extended to $0.05 \pm 0.06$, the whole region becomes compatible with oblique correction. 
\end{enumerate}

Another remarkable feature of the GM model is the appearance of the $H^{\pm}W^{\mp}Z$ vertex with a relatively greater strength than, for example a model with a single triplet. For $v_\chi = v_\xi$ this is the $H_5^{\pm}W^{\mp}Z$ vertex. Once the finite correction $\delta_0$ is incorporated, a non-vanishing $H_3^+W^-Z$ vertex also appears. To obtain these vertex strengths let us first construct the Goldstone mode\cite{Kundu:1995qb} in this case and define its orthogonal states to be $\psi_1^+$ and $\psi_2^+$; 
\begin{eqnarray}
    G^+ &=& \frac{v_\phi}{v} \phi^+ + \frac{2 v_\chi}{v} \chi^+ + \frac{2 v_\chi(1 + \delta_0)}{v} \xi^+ \nonumber \\
        &=& \phi^+\cos{\alpha} + \chi^+ \sin{\alpha}\cos{\beta} + \xi^+\sin{\alpha}\sin{\beta}  
\end{eqnarray}
\begin{eqnarray}
\psi_1^+ &=& -\phi^+\sin{\alpha} + \chi^+\cos{\alpha}\cos{\beta} + \xi^+\cos{\alpha}\sin{\beta}   \\
\psi_2^+ &=& -\chi^+\sin{\beta} + \xi^+\cos{\beta} 
\end{eqnarray} 
\noindent where, { $\tan{\alpha} = \frac{2v_\chi\sqrt{1 + (1 + \delta_0)^2}}{v_\phi} $} and $\tan{\beta} = 1 + \delta_0$.\newline
The singly charged physical states , $H_3^+$ and $H_5^+$ will be a linear combinations of  $\psi_1^+$ and $\psi_2^+$. The mixing angle depends on the loop corrections to the parameters of the potential, which are incalculable within the ambit of the model itself as discussed above. This angle, viz. $\gamma$ is a phenomenological input, yielding the charged scalar states as,
\begin{eqnarray}
    H_3^+ = \psi_1^+\cos{\gamma}  + \psi_2^+\sin{\gamma}  \\
    H_5^+ = -\psi_1^+\sin{\gamma}  + \psi_2^+\cos{\gamma} 
\end{eqnarray}
With this we can write the $H_3^+W^-Z$ and $H_5^+W^-Z$ vertex strengths as\cite{Kundu:1995qb},
\begin{eqnarray}
    V_{H_3^+W^-Z} &=& \frac{g^2}{\sqrt{2}\cos^2{\theta_W}}(C_3^\phi(\cos^2{\theta_W} - 1)\frac{v_\phi}{\sqrt{2}} + C_3^\chi(\cos^2{\theta_W} - 2){\sqrt{2}}{v_\chi} + C_3^\xi{\sqrt{2}}\cos^2{\theta_W}{v_\xi})
\end{eqnarray}

\noindent where, $C_3^\phi = -\sin{\alpha}\cos{\gamma}$, $C_3^\chi = (\cos{\alpha}\cos{\beta}\cos{\gamma} - \sin{\beta}\sin{\gamma}$) and $C_3^\xi = (\cos{\alpha}\sin{\beta}\cos{\gamma} + \cos{\beta}\sin{\gamma}$)

\begin{eqnarray}
    V_{H_5^+W^-Z} &=& \frac{g^2}{\sqrt{2}\cos^2{\theta_W}}(C_5^\phi(\cos^2{\theta_W} - 1)\frac{v_\phi}{\sqrt{2}} + C_5^\chi(\cos^2{\theta_W} - 2){\sqrt{2}}{v_\chi} + C_5^\xi{\sqrt{2}}\cos^2{\theta_W}{v_\xi})
\end{eqnarray}
\noindent where, $C_5^\phi = \sin{\alpha}\sin{\gamma}$, $C_5^\chi = -(\cos{\alpha}\cos{\beta}\sin{\gamma} + \sin{\beta}\cos{\gamma}$) and $C_5^\xi = (\cos{\beta}\cos{\gamma} - \cos{\alpha}\sin{\beta}\sin{\gamma} $)\newline
Let us define,
\begin{eqnarray}
    \kappa_{H_3^+W^-Z} = V_{H_3^+W^-Z}/M_W \\
    \kappa_{H_5^+W^-Z} = V_{H_5^+W^-Z}/M_W
\end{eqnarray}
 Note that  $\gamma = 0$ and $\beta = \frac{\pi}{4}$ in the limit $\delta_0 = 0$, in which case $V_{H_3^+W^-Z}$ vanishes identically. {We show below these two vertex strengths against $s_H$ and the colors indicate the values of $|\tan{\gamma}|$.}
\begin{figure}[!htb]
     \centering
     \begin{subfigure}[b]{0.45\textwidth}
         \centering
         \includegraphics[width=\textwidth]{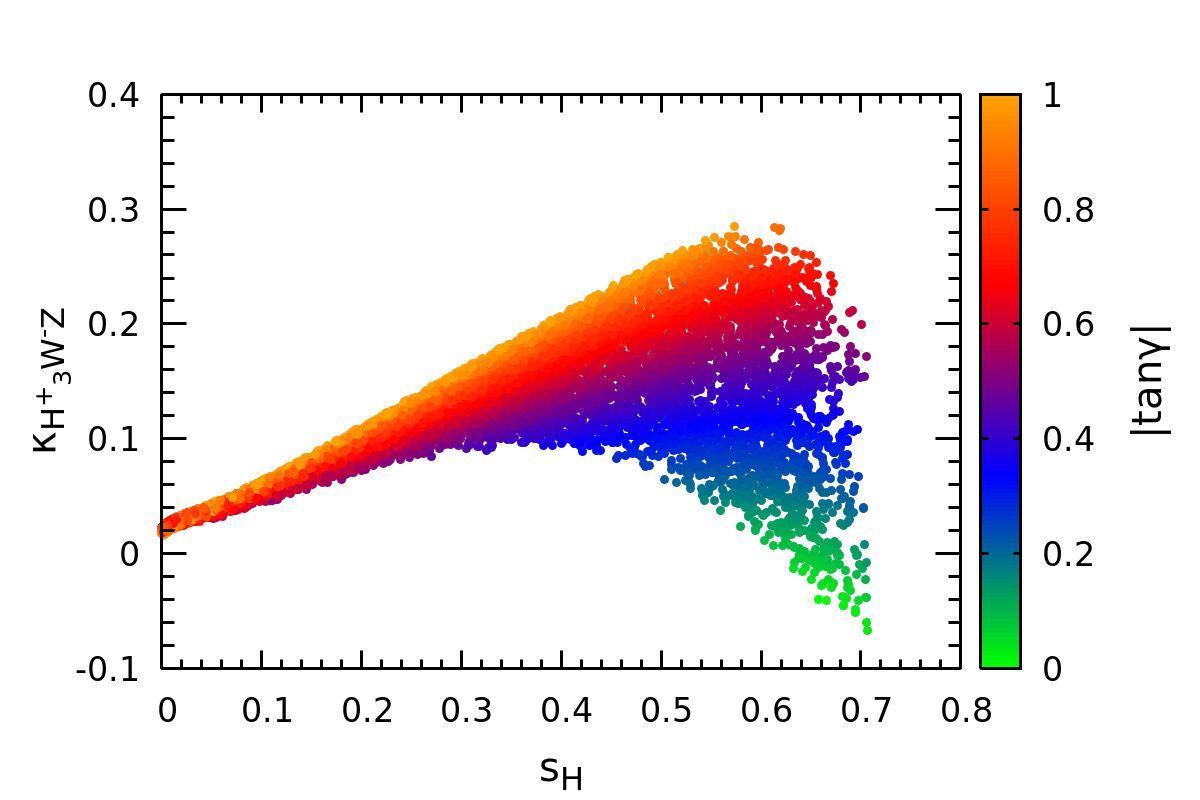}
         \caption{}
         \label{}
     \end{subfigure}
     \hspace{0.01\textwidth}
     \begin{subfigure}[b]{0.45\textwidth}
         \centering
         \includegraphics[width=\textwidth]{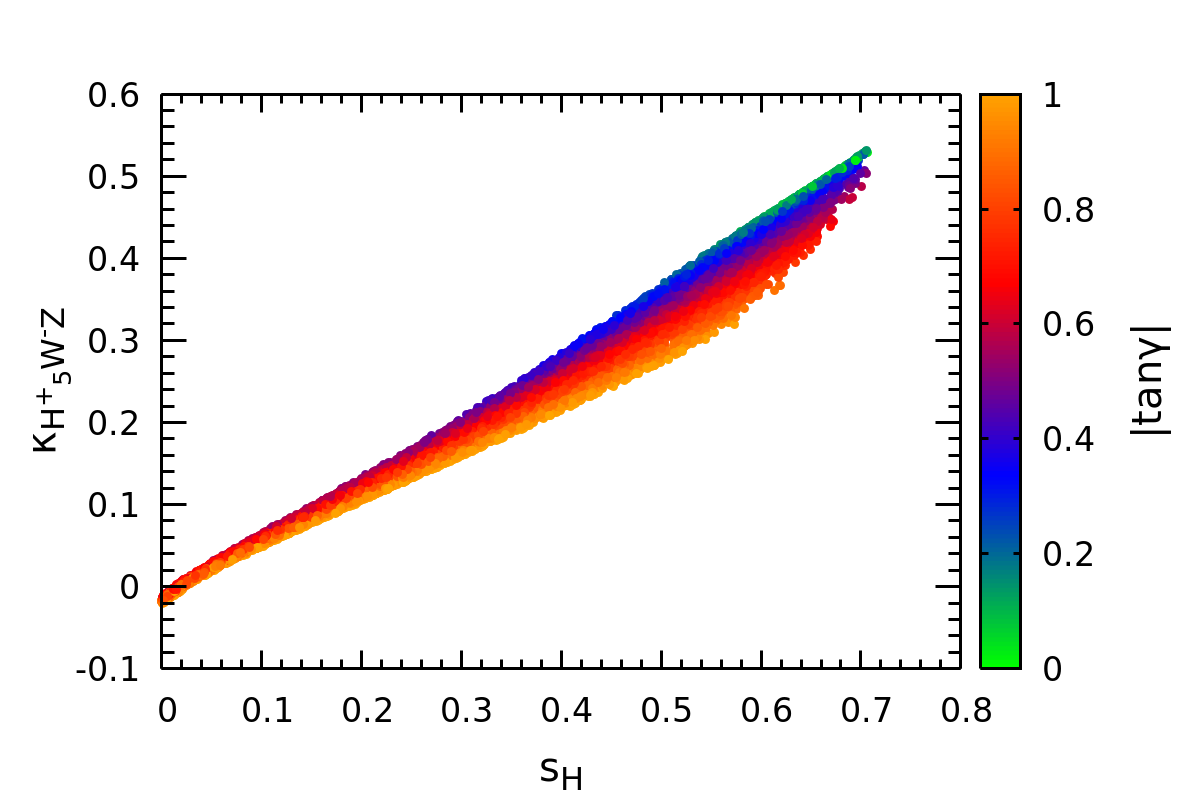}
         \caption{}
         \label{}
     \end{subfigure}
        \caption{The vertex strength modification factors $\kappa_{H_3^+W^-Z}$ and $\kappa_{H_5^+W^-Z}$ as function of $s_H$. The color axis represents the values of $|\tan{\gamma}|$.}
        \label{2f}
\end{figure}
It is seen in Fig. \ref{2f} that for high $\gamma$, $V_{H_3^+W^-Z}$ can be comparable to $H_5^+W^-Z$, while for small $\gamma$, $V_{H_5^+W^-Z}$, is always dominant. Obviously $\gamma$ can be large only when the UV completion is strongly coupled. As seen from Fig.   \ref{1f}, this can happen for large $\delta_0$ which in turn implies small values of $s_H$. Therefore, the $H_3^+W^-Z$ interaction strength is likely to be of less phenomenological consequence if one has to believe in the CDF claims on $M_{W}$, some likely exception coming for morderate values of $\delta_0$, $s_H$.  \newline
This alerts one to the alternative possibility of allowing $v_\chi \ne v_\xi$ at the tree level itself, fitting the CDF claim in terms of this splitting of vevs. Note that the parameters in the potential become phenomenological inputs in this approach. While we are deliberately refraining from proposing theoretical mechanisms behind this, we remind the reader that an `{\it approximate}' custodial symmetry is not at all an unfamiliar proposition.{When the custodial SU(2) is broken, one needs to go beyond the form of scalar potential in equation \eqref{11}. {One of the ways to do so is to use the most general potential under $SU(2)_{L}\times U(1)_Y$, following} reference\cite{Keeshan}}
{

\begin{eqnarray}
	V(\phi, \chi, \xi) &=& \tilde \mu_2^2 \phi^{\dagger} \phi + \tilde \mu_3^{\prime 2} \chi^{\dagger} \chi 
		+ \frac{\tilde \mu_3^2}{2} \xi^{\dagger} \xi  + \tilde \lambda_1 (\phi^{\dagger} \phi)^2 
		+ \tilde \lambda_2 | \tilde \chi^{\dagger} \chi |^2 
		+ \tilde \lambda_3 (\phi^{\dagger} \tau^a \phi) (\chi^{\dagger} t^a \chi)  \nonumber \\
	&& + \left[ \tilde \lambda_4 (\tilde \phi^{\dagger} \tau^a \phi)(\chi^{\dagger} t^a \xi) 
			+ {\rm h.c.} \right] + \tilde \lambda_5 (\phi^{\dagger} \phi)(\chi^{\dagger} \chi)
		+ \tilde \lambda_6 (\phi^{\dagger} \phi)(\xi^{\dagger} \xi) + \tilde \lambda_7 (\chi^{\dagger} \chi)^2 \nonumber \\
	&&	+ \tilde \lambda_8 (\xi^{\dagger} \xi)^2 + \tilde \lambda_9 | \chi^{\dagger} \xi |^2
		+ \tilde \lambda_{10} (\chi^{\dagger} \chi) (\xi^{\dagger} \xi) - \frac{1}{2} \left[ \tilde M_1^{\prime} \phi^{\dagger} \Delta_2 \tilde \phi + {\rm h.c.} \right] \nonumber \\
	&& + \frac{\tilde M_1}{\sqrt{2}} \phi^{\dagger} \Delta_0 \phi - 6 \tilde M_2 \chi^{\dagger} \overline \Delta_0 \chi
   \label{eq:potential3}
\end{eqnarray}
where,
\begin{eqnarray}
	\Delta_2 &\equiv& \sqrt{2} \tau^a U_{ai} \chi_i 
		= \left( \begin{array}{cc} \chi^+/\sqrt{2} & -\chi^{++} \\ 
			\chi^0 & -\chi^+/\sqrt{2} \end{array} \right), \nonumber \\
	\Delta_0 &\equiv& \sqrt{2} \tau^a U_{ai} \xi_i
		= \left( \begin{array}{cc} \xi^0/\sqrt{2} & -\xi^+ \\
			-\xi^{+*} & -\xi^0/\sqrt{2} \end{array} \right), \nonumber \\
	\overline \Delta_0 &\equiv& -t^a U_{ai} \xi_i
		= \left( \begin{array}{ccc} -\xi^0 & \xi^+ & 0 \\
			\xi^{+*} & 0 & \xi^+ \\ 0 & \xi^{+*} & \xi^0 \end{array} \right).
\end{eqnarray}
{If one minimises the above potential , neglecting for simplicity any CP-violating phase in $\Tilde{\lambda_4}$ and $\Tilde{M_1}^\prime$, one obtains the following conditions by setting the tadpoles to zero,}

\begin{eqnarray}
    0 = \frac{\partial V}{\partial v_{\phi}} &=&  v_{\phi}     [\tilde \mu_2^2 
		+ \tilde \lambda_1 v_{\phi}^2 
		+ \frac{\tilde \lambda_3}{2} v_{\chi}^2
		+ \sqrt{2} \tilde \lambda_4 v_{\chi} v_{\xi} \nonumber \\ 
	&&    + \tilde \lambda_5  v_{\chi}^2
		+ \tilde \lambda_6 v_{\xi}^2 
		- \tilde M_1^{\prime} v_{\chi} 
		- \frac{\tilde M_1}{2} v_{\xi} ], 
	\label{eq:potential5} \\
	0 = \frac{\partial V}{\partial v_{\chi}} &=& 2 \tilde \mu_3^{\prime 2} v_{\chi} 
	+ \frac{\tilde \lambda_3}{2} v_{\phi}^2 v_{\chi} 
	+ \frac{\tilde \lambda_4}{\sqrt{2}} v_{\phi}^2 v_{\xi}
	+ \tilde \lambda_5 v_{\phi}^2 v_{\chi} \nonumber \\
	&& + 4 \tilde \lambda_7 v_{\chi}^3 
	+ 2 \tilde \lambda_{10} v_{\chi} v_{\xi}^2 	 
	- \frac{\tilde M_1^{\prime}}{2} v_{\phi}^2 \nonumber \\
	&& - 12 \tilde M_2 v_{\chi} v_{\xi}, 
	\label{eq:potential6} \\
	0 = \frac{\partial V}{\partial v_{\xi}} &=& \tilde \mu_3^2 v_{\xi} 
	+ \frac{\tilde \lambda_4}{\sqrt{2}} v_{\phi}^2 v_{\chi} 
	+ \tilde \lambda_6 v_{\phi}^2 v_{\xi} 
	+ 4 \tilde \lambda_8 v_{\xi}^3 \nonumber \\
	&& + 2 \tilde \lambda_{10} v_{\chi}^2 v_{\xi} 
	- \frac{\tilde M_1}{4} v_{\phi}^2  
	- 6 \tilde M_2 v_{\chi}^2.
	\label{eq:potential7}
\end{eqnarray}
which reveals that ${ v_\chi}$ $\ne$ ${ v_\xi}$ in general. The ranges of various parameters as used in our analysis are indicated in Table \ref{t3}. $\Tilde \lambda_1$ is tuned such that one of the three CP-even Higgs masses is always fixed at 125 GeV\cite{Keeshan}. ${\tilde M}_5^{\pm\pm}$ denotes the mass of the doubly charged state. ${\Tilde M}_5^{\pm}$ denotes the mass of the singly charged scalar that has the least superposition with $\phi^+$ and has the smallest mass splitting with ${\Tilde M}_5^{++}$. ${\Tilde M}_3^{\pm}$ is the mass of the other singly charged scalar. The masses are obtained by exactly diagonalizing the resulting mass matrix from the potential given in \eqref{eq:potential3}. ${\Tilde M}_5^{++}$ is given by,
\begin{eqnarray}
	{\Tilde M}^{++2}_{5} &=& 4 \tilde \lambda_2 v_{\chi}^2 
		- \frac{\tilde \lambda_3 v_{\phi}^2}{2} 
		- \frac{\tilde \lambda_4 v_{\phi}^2 v_{\xi}}{2 \sqrt{2} v_{\chi}} 
		+ \frac{\tilde M_1^{\prime}}{4 v_{\chi}} v_\phi^2 
		+ 12 \tilde M_2 v_{\xi}.
	\label{eq:potential8}
\end{eqnarray}
As is evident from the entries in the table, a combination of physical mass and gauge basis parameters have been used as our inputs. 
}

\begin{table}[]

    \centering
    \begin{tabular}{|c|c|c|}
 \hline
 $\tilde{M_1}$ & 100  GeV \\
 \hline
 $\tilde{M_2}$ & 10 GeV  \\ 
 \hline
 ${\Tilde M}_5^{++}$ & 200-1000 GeV  \\ 
 \hline
 ${\Tilde M}_5^{+},{\Tilde M}_3^{+}$ & $>$ ${\Tilde M}_5^{++}$  \\ 
 \hline
 $\frac{v_\phi}{v}$ & [0,1] \\
 \hline
 $\frac{v_\chi}{v}$ & [0,$\frac{1}{2}{\sqrt{1 - (\frac{v_\phi}{v})^2}}$] \\
 \hline
    \end{tabular}
    \caption{ Values of various parameters used in our scan to get Fig. \ref{3f} and \ref{4f}.  {The quartic couplings (other tan $\lambda_1$) have been scanned in the range [$-4\pi,4\pi$] in order to reproduce the masses.}}
    \label{t3}
\end{table}

$v_\chi \ne v_\xi$ at tree level frees one from the restriction imposed by equation \eqref{10}. As a result, an allowed  region
of the parameter space opens up, where the incompatibility between the existing 
$M_Z$ and the CDF measurement $M_{W}$  goes away. In this case

\begin{eqnarray}
    M_W^2 &=& \frac{g^2}{4}(v_\phi^2 + 4v_\chi^2 + 4v_\xi^2) \\ \nonumber
         &=& \frac{g^2 }{4}(v_\phi^2 + 8v_\chi^2) + g^2(v_\xi^2 - v_\chi^2)
\end{eqnarray}

The custodial symmetry breaking effect should be small as reflected in the $\rho$ parameter. We can equate the first term to $M_{W,old}$ (Since it is consistent, with measured $M_Z$ as per the discussion in section 2) and expect the second term to generate the shift required to match the CDF claim. Hence,
\begin{equation}
   \Delta M_W = \frac{4(v_\xi^2 - v_\chi^2)}{2v^2}{M_{W,old}} 
\end{equation}
Fig.  \ref{3fa} and \ref{3fb} exhibit the allowed regions corresponding to $M_{W,old}$ fixed according to the ATLAS result and global average respectively.
The regions are specified by the co-ordinates $v_\chi$ and $\Delta v_\chi$ where 
$\Delta v_\chi = v_\xi - v_\chi$. $\Delta v_\chi < 0$ is not used in the scan,
since such a hierarchy does not serve to raise the W-mass to match the CDF claim.
\footnote{$M_{W,old}$ equated with $LEP + Tevatron$ measurement continues to be consistent with the CDF claim in anyway.}
{The scan over parameters in each case is once more subject to the SM-like Higgs mass being in the right region, quartic couplings being perturbative and vacuum stability being satisfied. The results corresponding to the two aforementioned benchmarks of  $M_{W,old}$ are shown in Fig.  \ref{3f}. {  For illustration, we also show $\Delta M_W$ against $v_\chi$ in Fig. \ref{4e}}
\vspace{0.2cm}

{The points to note further are as follows:}

\begin{itemize}
\item { The generalized Georgi-Machacek scenario with custodial symmetry broken at tree level has a multidimensional parameter space, since its potential has
3 mass terms, 10 quartic interactions, and 3 trilinear terms involving the doublet and triplet scalars. It is not our purpose here to scan the entire parameter space. We instead demonstrate its consistency with the CDF result  in terms of some illustrative situations with $v_\chi \ne v_\xi$, with random scan over 
some parameters. The EWSB conditions are satisfied for each point in the parameter space displayed in Fig.  \ref{3f} in $\Delta v_\chi = v_\xi - v_\chi$ vs $v_\chi$ plane.} 


\item { The lower limits on $v_\chi$ arise when one has non-zero values of the trilinear term parameters
$\Tilde{M_1}^\prime, \Tilde{M_2}$ and both the fields $\phi$ and $\xi$ aquired a nonzero vev. The parameters are further constrained such that the SM-like scalar mass has the right value, the doubly charged scalar mass, ${\Tilde M}_5^{++}$ lies between 200-1000 GeV and ${\Tilde M}_5^{++} < {\Tilde M}_3^+, {\Tilde M}_5^+$.} 

\item {The maximum value of $\Delta v_\chi$ in each case comes from the requirement of satisfying
the $2\sigma$ limit on the $\rho$-parameter. The small minimum value for each $v_\chi$ answers to the smallest split in vev required to reconcile $M_W$ with $M_Z$ and $\rho$.}

\item  { The doubly charged scalar mass used for illustration is between 200-1000 GeV, and for such a combination of  
parameters  that it decays only into the channel $H_5^{++}\rightarrow$ $W^+ W^+$.}
The limits on the parameter space
from  the  data on the  H$^{++}$ search in  vector-boson fusion channel \cite{ref38}\cite{ref42} are used for all points generated in the random scan, consistently with the above benchmarking. It may also be noted that, for the points in the parameter space used for our illustration, the constraints from VBF search mostly turns out to be the strongest. Flavour constraints such as those from $b \rightarrow s \gamma$ and $B_s \rightarrow \mu^+ \mu^-$, and also that from the S-parameter are weaker\cite{rb2022,ref38}.
\newline
\end{itemize}
{As can be seen from Fig.  \ref{3f}, one obtains rather non-negligible regions satisfying all constraints, for
the illustrative benchmarks. It is thus possible to reconcile the GM scenario  with the CDF claim, so long as a  broken custodial SU(2) is conceded, without compromising on the limits from the $\rho$-parameter.}

     \begin{figure}[!htb]
     \centering
     \begin{subfigure}[b]{0.480\textwidth}
         \centering
         \includegraphics[width=\textwidth]{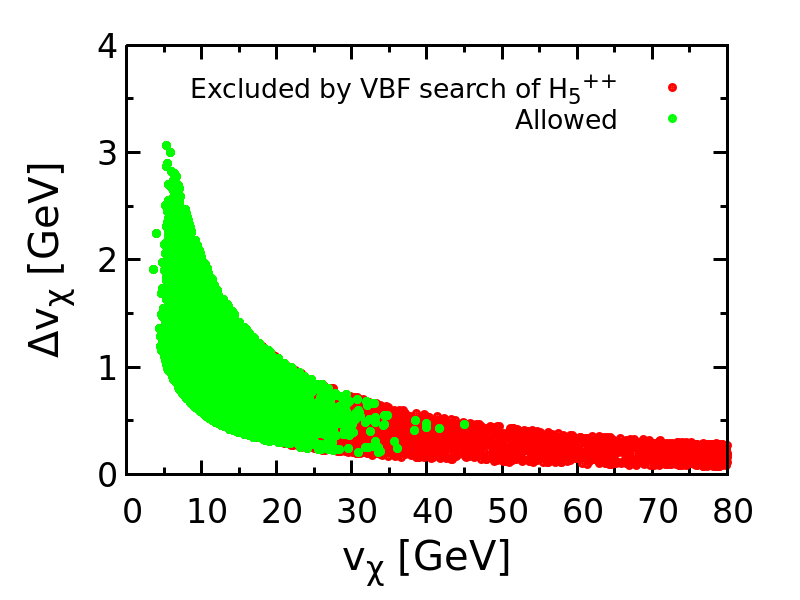}
         \caption{}
         \label{3fa}
     \end{subfigure}
     \hspace{0.01\textwidth}
     \begin{subfigure}[b]{0.480\textwidth}
         \centering
          \includegraphics[width=\textwidth]{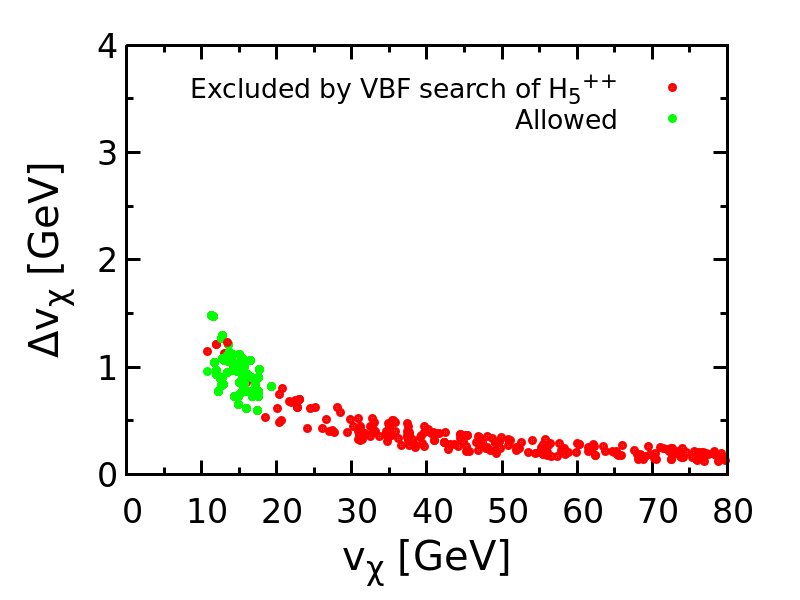}
         \caption{}
        \label{3fb}
     \end{subfigure}
    
        \caption{(a) and (b): Regions in the $v_\chi$ - $\Delta v_\chi$ plane which satisfy theoretical as well as $\rho$ parameter constraints but are excluded by VBF data of $H_5^{++}$ search at 35.9 $fb^{-1}$ are shown in red, and those that satisfy the VBF constraints are in green. (a) uses the ATLAS measurement for $M_{w,old}$, while (b) uses the global fit, both at $2\sigma$ levels. \newline}
        \label{3f}
\end{figure}

\begin{figure}[!htb]
     \centering
         \includegraphics[width=0.6\textwidth]{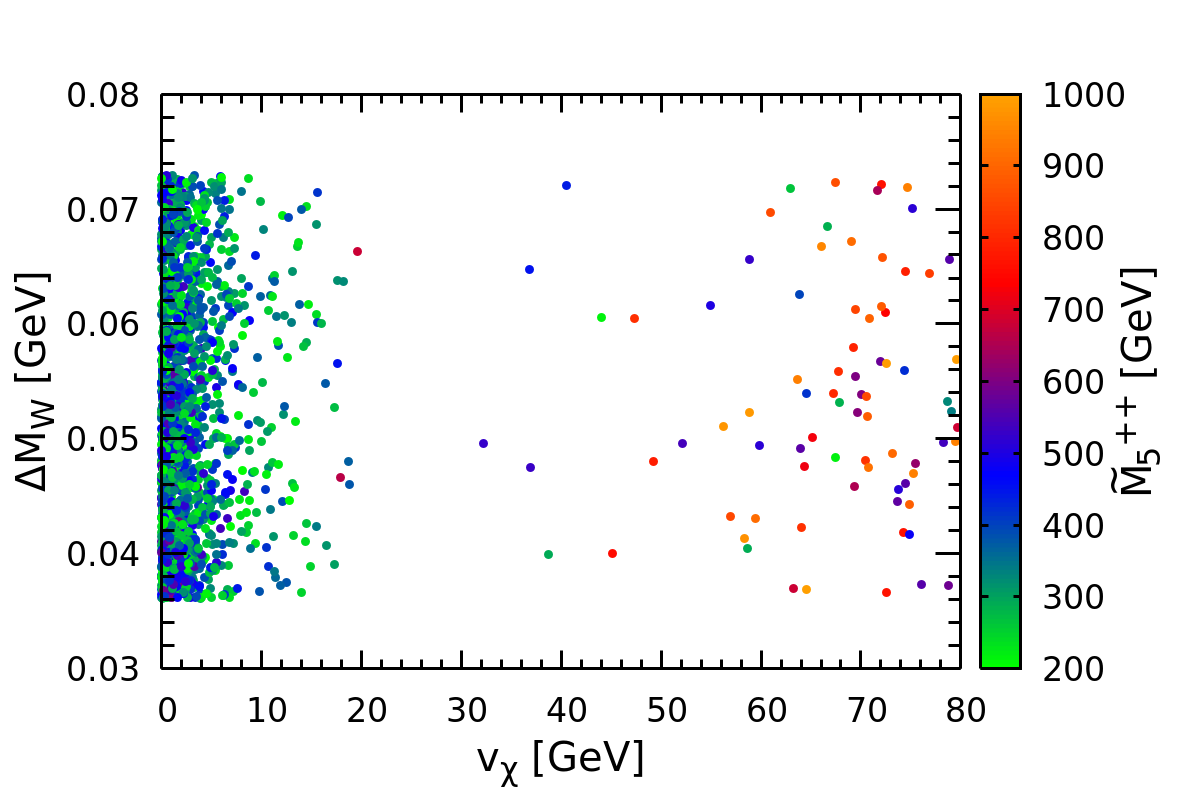}
        \caption{ {$\Delta M_W$ required to satisfy the CDF claim at 2$\sigma$ level as a function of complex triplet vev $v_\chi$. The color axis represents the mass of the doubly charged Higgs ${\Tilde M}_5^{++}$}}
        \label{4e}
\end{figure}

{ The $H^{\pm}W^{\mp}Z$ vertex strengths in this case are shown in Fig.  \ref{4f}. Here we define $v_{triplet} = \sqrt{4v_\chi^2 + 4v_\xi^2}$.
Note that the vertex like $H_5^{++}W^{-}W^{-}$ also appears with higher strength in GM like scenario with respect to the models with a single triplet. But irrespective of the presence of custodial symmetry this vertex strength is always proportional to $v_\chi$ and since for the custodial symmetric case consistency with oblique correction requires triplet vev to be high, this vertex strength is always high i.e $ 0.6 \le \frac{V_{H_5^{++}W^{-}W^{-}}}{\sqrt{2}gM_W} \le 0.7$. When the custodial symmetry is broken at tree level the vertex strength has a larger range to vary i.e $0.1 \le \frac{V_{H_5^{++}W^{-}W^{-}}}{\sqrt{2}gM_W} \le 0.9$ (These limits are given without collider data). On the other hand, custodial symmetry has a more significant effect on the vertex strength of $H^\pm W^\mp Z$ and hence here we are interested about this vertex only.}
\begin{figure}[!htb]
     \centering
     \begin{subfigure}[b]{0.45\textwidth}
         \centering
         \includegraphics[width=\textwidth]{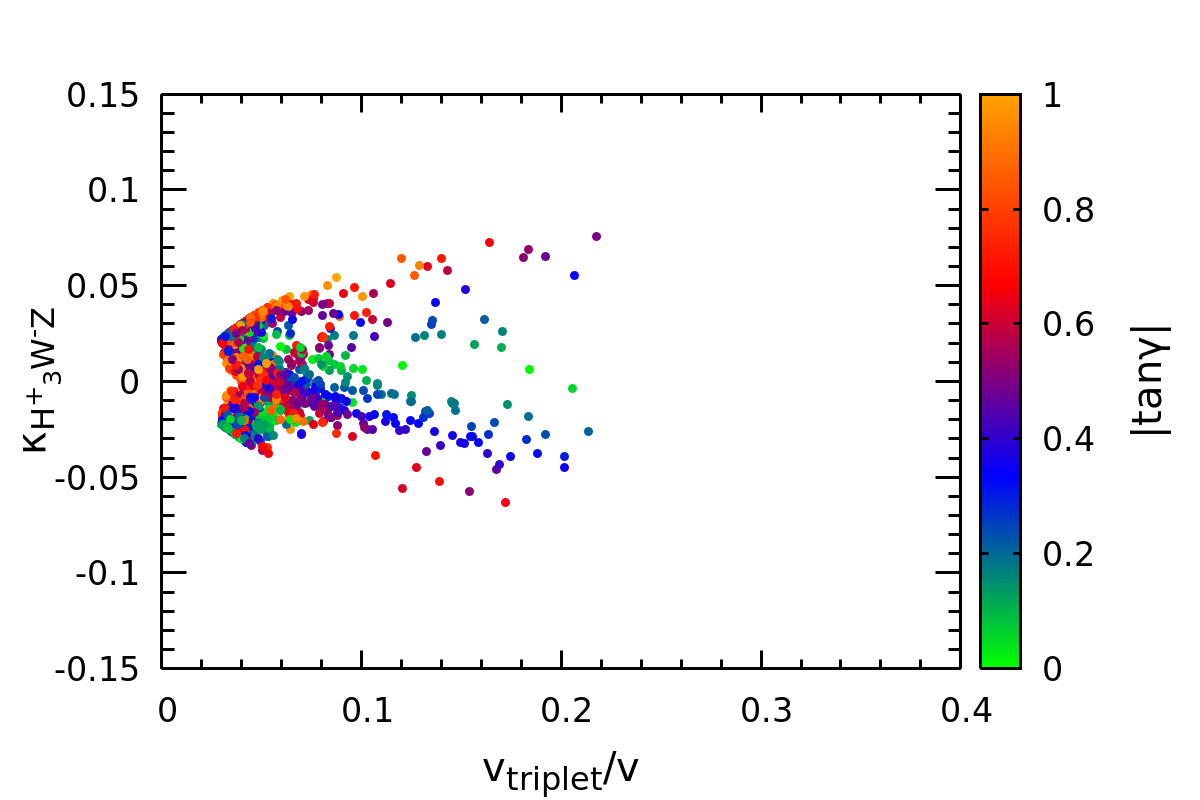}
         \caption{}
         \label{}
     \end{subfigure}
     \hspace{0.01\textwidth}
     \begin{subfigure}[b]{0.45\textwidth}
         \centering
         \includegraphics[width=\textwidth]{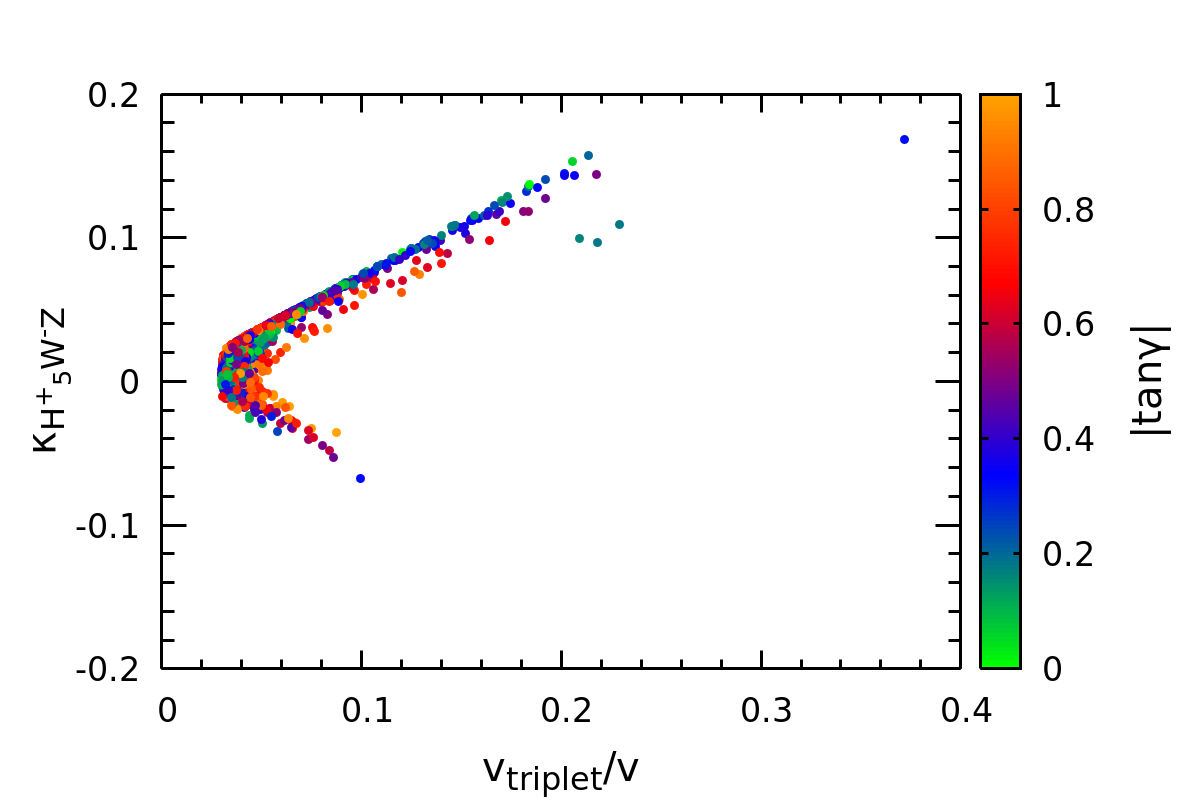}
         \caption{}
         \label{}
     \end{subfigure}
        \caption{ The vertex strength modification factors $\kappa_{H_3^{\pm}W^{\mp}Z}$ and $\kappa_{H_5^{\pm}W^{\mp}Z}$ as function of $v_{triplet}/v$. The color axis represents the modulus of $\tan\gamma$.}
        \label{4f}
\end{figure}
\subsection{A single scalar triplet}
For the sake of completeness, we include below a scenario with a $Y=2$, complex triplet $\Delta$ in addition to the SM doublet $\Phi = (\phi^+,\phi^0)$. { It should be noted that the CDF-claim has also been analyzed with the addition of just one real triplet to the usual Higgs doublet \cite{realt}. We however, discuss the complex triplet case only, because of its relevance in neutrino mass generation.} The scalar potential has the following form, 
\begin{eqnarray}
    V(\Phi,\Delta) &=& -m_\Phi^2 \Phi^\dagger \Phi + M^2 Tr \Delta^\dagger \Delta \nonumber \\
                   &+& (\mu \Phi^T i \sigma^2 \Delta^\dagger \Phi + h.c) + \frac{\lambda}{4}(\Phi^\dagger \Phi)^2 \nonumber \\
                   &+& \lambda_1 \Phi^\dagger \Phi Tr\Delta^\dagger \Delta + \lambda_2 (Tr\Delta^\dagger \Delta)^2 \nonumber \\
                   &+& \lambda_3 Tr(\Delta^\dagger \Delta)^2 + \lambda_4 \Phi^\dagger \Delta \Delta^\dagger \Phi
\end{eqnarray}
with 
\begin{equation}
    \Delta = \begin{pmatrix}
    	\frac{\Delta^{+}}{\sqrt{2}} & \Delta^{++}\\
		\Delta^0 & \frac{-\Delta^{+}}{\sqrt{2}}
		\end{pmatrix}
\end{equation}
{ The ranges of our scanned over parameters are given in Table \ref{t4}. $\lambda$ is tuned such that one of the two CP-even Higgs masses is always fixed at 125 GeV\cite{ref32}. $M_H^{++}$ is the mass of the doubly charged scalar. The mass of the doubly charged and the singly charged scalar are given by ,
\begin{eqnarray}
    {M_H^{++}}^{2} = \frac{\mu v_d^2}{\sqrt{2}{v_t}} - \frac{\lambda_4}{4}v_d^2 - \lambda_3 v_t^2 \\
    {M_H^{+}}^{2} = \frac{(2\sqrt{2}\mu - \lambda_4 v_t) }{4 v_t}(v_d^2 + 2 v_t^2)
\end{eqnarray}
}
\begin{table}[]

    \centering
    \begin{tabular}{|c|c|c|}
    \hline
        $\lambda_1$ - $\lambda_4$ & $[-4\pi,4\pi]$  \\
        \hline
         $M_H^{++}$ & $> 400$ GeV \\
         \hline
         $\frac{v_t}{v}$ & $[0,0.1]$ \\
         \hline
    \end{tabular}
    \caption{\centering  Range of parameters scanned to produce Fig. \ref{5f}}
    \label{t4}
\end{table}
The neutral component of $\Delta$ is instrumental in generating
neutrino masses when one allows gauge-invariant $\Delta L = 2$ Yukawa couplings with the SM leptons\cite{ref29,ref32,ref33}. 
Such couplings can also in principle generate new collider phenomenology, the most striking instance
being  $H^{++}$ giving rise to $\ell^+ \ell^+$ or $W^+ W^+$ pairs, where $H^{++} = \Delta^{++}$\newline 
Simultaneous nonzero values of $\langle\phi^0\rangle = v_d/\sqrt{2}$ and 
$\langle\Delta^0\rangle = v_t/\sqrt{2}$, automatically shifts the value of $\rho$ away from 1. Therefore, it is imperative for $v_t$ to be  small enough not to overshoot  of the allowed interval of $\rho$, as derived from  the oblique electroweak parameter T \cite{ref34}.
Electroweak precision data, mainly the T-parameter, restrict the mass splitting between the doubly-and 
singly-charged scalars. In this study we have selected our benchmarks satisfying 
$|M_H^{++} - M_H^+| \le 40$ GeV, which is
consistent with the T-parameter. We have calculated the oblique parameters following reference \cite{ref32,tript}  The triplet couplings to leptons are taken to be diagonal.
Among collider constraints,  the production of doubly charged Higgs via vector boson fusion (VBF) does not appear to be significant as the triplet vev is already small from the restriction imposed by the $\rho$-parameter. 
Instead, the Drell Yan(DY) pair production of doubly charged Higgs and the dilepton decay of the same puts a major constraint on the parameter space, although such constraints depend on the decay
branching ratios of $H^{\pm\pm}$. The ATLAS search  for DY production of $H^{++}$ pair and its subsequent decay to diboson channel restricts $M_H^{++} \ge 220$ GeV. The constraint on a leptonically decaying 
 doubly charged scalar becomes stringent in the low triplet vev limit. For $v_t = 10^{-4}$ GeV, this constraint requires $M_H^{++} \ge 450$ GeV \cite{ref32}
 Once more, our results correspond to benchmarks that appropriately satisfy all these constraints. \newline
In Fig. \ref{5f}, we show the allowed region at $2\sigma$ level in the $v_t - v_d$ parameter space of a scenario that yields the SM-like scalar mass in its allowed interval around $125$ GeV. The allowed region is further restricted by the new measurement of $M_W$ and the existing value of $M_Z$, consistently with the requirements of vacuum stability and perturbative unitarity\cite{ref32,ref33}. 
Fig.  \ref{5f}  thus reflects the capability of a single-triplet scenario
in simultaneously explaining $M_W$ as announced by CDF,  $M_Z$, and also the $\rho$-parameter. Also, this scenario continues to be useful in neutrino mass generation, since the low-$v_t$ ranges are unaffected.

\begin{figure}[!htb]
         \centering
         \includegraphics[width=0.6\textwidth]{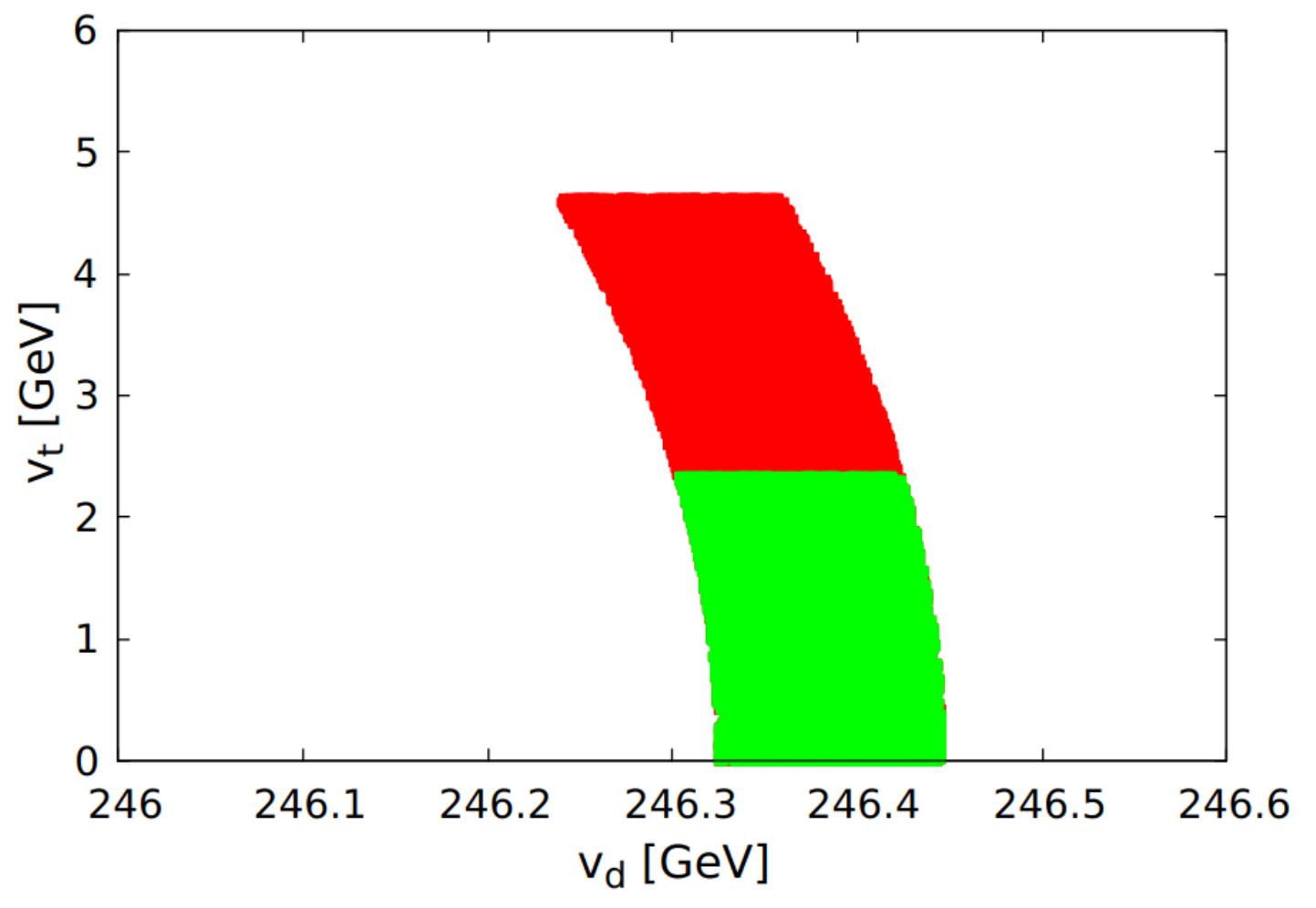}
        \caption{ The red region is compatible with $M_W$ as measured by CDF but excluded by electroweak precision tests. The green region is allowed by both.}
        \label{5f}
\end{figure}

 \section{Summary and conclusions}
 We have shown that, if both the value of $M_W$ announced by CDF and  the extant value of 
 $M_Z$ continue to remain within their respective $2\sigma$ intervals, then any EWSB sector 
 which  theoretically predicts $\rho = 1$ at the tree-level becomes inconsistent. This is true if one accepts (a) the earlier value of $M_W$ obtained from global fits, or (b) the
current value from the ATLAS collaboration. On the other hand, The new $M_W$ and
the extant $M_Z$ are consistent within $1\sigma$ with $\rho = 1$ if the earlier $M_W$ obtained
by using just the {\em LEP + Tevatron} data. 

Keeping this in view, we have considered the Georgi-Machacek scenario comprising a complex
$Y = 2$  scalar triplet, a real $Y = 0$ triplet, and the SM Higgs doublet, which
retains $\rho = 1$ at the tree-level. This {\it prima facie} disqualifies this 
model at the tree level for two of the three reference points of  $M_{W,old}$. However we show, that the CDF claim together with the experimental value of $\rho$ parameter can be satisfied if one includes the finite corrections to $M_w$ which is not calculable from the first principles unless one knows the UV completion of the GM scenario. It is also shown that this finite correction should be small for higher triplet vev around 30 GeV or above. On the other hand, a low triplet vev may allow relatively large correction. We have also explored the possibility of an explicit small splitting between the two triplet vevs at the tree level itself and have identified the allowed region in $\Delta v_\chi - v_\chi$ plane. In context of colliders, both of these scenarios are interesting due to the presence of nonzero $H^+W^-Z$ vertices. It has been shown that as long as the tree level scalar potential respects custodial $SU(2)$, only $H_5^+W^-Z$ vertex can be significant. On the other hand, for either of the two schemes outlined to accomodate the the CDF claims the $H_3^+W^-Z$ vertex strenghths can be  not only significant but also sometimes comparable to the $H_5^+W^-Z$ vertex strength. 

For the sake of completeness we take into account, the simplest extension of the EWSB sector with $\rho \ne 1$, namely a scenario
with a scalar doublet  and a singlet complex $Y = 2$  scalar triplet. It turns out to be 
compatible with the new $M_W$, the existing $M_Z$ and the experimental range of 
$\rho$, for all the three ways of standardizing   $M_{W,old}$ mentioned above once the oblique parameters are considered.

Before we conclude, some comments are in order regarding the additional, $\Delta L = 2$ Yukawa couplings involving the $Y = 2$ triplet and the leptons. In principle, these couplings may be subject to contraints from flavour-changing neutral current(FCNC) process such as $\mu \rightarrow e \gamma$ and $\mu \rightarrow 3e$. Moreover such couplings may contribute to muon decay and thus tend to alter the Fermi coupling contant extracted thereform. These and similar kinds of issues are addressed by the fact that the ($\Delta L = 2$) Yukawa coupling strengths relevant in the parameter region in Fig.  \ref{5f} corresponds to $ v_t \le 10^{-11}$. Such values are too small to effect either $G_F$ or FCNC rates, even if there remain flavour-changing $\Delta L = 2$ Yukawa interactions. On the contrary, Yukawa couplings which can affect the above processes will require non-vanishing triplet vev $\le 10^{-6}$ GeV (from neutrino mass limits)\cite{ref32}\cite{Chun:2003ej}. Such a vev has no perceptible consequence on the W-boson mass, and is thus not relevant for the present discussion.
\newline

{\bf Acknowledgments}: We thank Jayita Lahiri and Tousik Samui for helpful comments and Ritesh K Singh for technical help. The work of
 RG has been supported by a fellowship awarded  by University Grants Commission, India, while US acknowledges support from Department of Atomic Energy, Government of India for a research grant associated with the Raja Ramanna Fellowship.
 \newline

\end{document}